\documentclass{ws-mpla}
\usepackage[super]{cite}
\usepackage{graphicx}

\newcommand{\pv}{\langle\phi\rangle}
\newcommand{\hv}{\langle H_u^0\rangle}

\begin{document}

\markboth{Bibhushan Shakya}
{Sterile Neutrino Dark Matter from Freeze-In}

\catchline{}{}{}{}{}

\title{STERILE NEUTRINO DARK MATTER FROM FREEZE-IN}

\author{\footnotesize BIBHUSHAN SHAKYA}

\address{Michigan Center for Theoretical Physics\\
University of Michigan, Ann Arbor, MI 48109, USA\\
bshakya@umich.edu}

\maketitle


\begin{abstract}
A sterile neutrino is a well-motivated and widely studied dark matter candidate. The most straightforward realization of sterile neutrino dark matter, through the Dodelson-Widrow mechanism, is now ruled out by a combination of X-ray and Lyman-$\alpha$ measurements. An alternative production mechanism that is becoming increasingly popular in the literature is the freeze-in mechanism, involving frameworks where a feeble coupling to a particle -- usually a scalar beyond the Standard Model -- in the thermal bath results in a gradual accumulation of the sterile neutrino dark matter abundance. This article reviews the various motivations for realizing such frameworks in the literature, their common characteristic features, and phenomenological signatures. 

\end{abstract}


\section{Introduction and Motivation}	

The identity of dark matter (DM) remains one of the greatest mysteries in physics. The WIMP miracle, coupled with the hierarchy problem, has long provided a compelling argument that dark matter consists of weakly interacting massive particles (WIMPs). However, a wide variety of indirect and direct detection experiments have reached the sensitivity to comprehensively probe typical WIMP interactions and are yet to yield any conclusive signals, placing the WIMP paradigm under significant tension. This provides added incentive to explore dark matter candidates beyond the WIMP. 

A well-motivated dark matter candidate emerges from the neutrino sector, where the observations of solar and atmospheric neutrino oscillations provide clear evidence for physics beyond the Standard Model (SM). The established framework for obtaining the tiny neutrino masses necessitated by the observed oscillations is the seesaw mechanism, which requires an extension of the SM to include right-handed, sterile neutrinos. While the mass scale of these sterile neutrinos is relatively unconstrained and can lie anywhere from the eV to the GUT scale, a keV scale sterile neutrino has a lifetime longer than the age of the Universe as well as the correct relic abundance to explain dark matter (Section \ref{sec:sndm}), making it an interesting candidate worthy of further scrutiny.  

The traditionally studied sterile neutrino dark matter candidate has a keV scale mass and is produced via its mixing with the active neutrinos through the Dodelson-Widrow (DW) mechanism \cite{Dodelson:1993je}, resulting in a warm dark matter candidate. This incarnation has been extensively studied in the framework of the Neutrino Minimal Standard Model ($\nu$MSM) \cite{Asaka:2005an,Asaka:2005pn,Asaka:2006nq}. More recently, a 7 keV sterile neutrino dark matter candidate has received significant interest due to the observation of an unidentified X-ray line at 3.5 keV in the stacked X-ray spectra of 73 galaxy clusters measured by XMM-Newton \cite{Bulbul:2014sua} and in the X-ray spectra of the Andromeda galaxy and the Perseus galaxy cluster \cite{Boyarsky:2014jta}. However, a combination of X-ray \cite{Boyarsky:2006fg,Boyarsky:2006ag, Boyarsky:2005us,Boyarsky:2007ay,Boyarsky:2007ge} and Lyman-$\alpha$ measurements \cite{Seljak:2006qw, Asaka:2006nq, Boyarsky:2008xj} now rule out the prospect of a sterile neutrino produced from DW accounting for all of dark matter (Section \ref{sec:sndm}). Nevertheless, given the strong motivation for the existence of sterile neutrinos and its appeal as a dark matter candidate, other scenarios that circumvent the combined X-ray and Lyman-$\alpha$ bounds and successfully realize it as a dark matter candidate have been explored. These include resonant production \cite{Shi:1998km}, extended frameworks that realize thermal freeze-out followed by entropy dilution\cite{Boyarsky:2008mt,Nemevsek:2012cd,Asaka:2006ek,Asaka:2006nq,Bezrukov:2009th,Patwardhan:2015kga}, interactions of light vector bosons \cite{Shuve:2014doa,Khalil:2008kp}, pion decay \cite{Lello:2014yha}, and neutrino mixing with a hidden dark sector \cite{Falkowski:2011xh}. 

Another option, pursued by several groups, is to produce the sterile neutrino dark matter relic density through the freeze-in mechanism \cite{Chung:1998rq, Hall:2009bx}, which carries the dual virtues of producing a colder sterile neutrino population compared to DW and not relying on any mixing with the active neutrinos for production\footnote{ It should be clarified that production through DW is, technically, also a freeze-in mechanism, as the sterile neutrino population does not equilibrate with the thermal bath. For the purpose of this article, freeze-in will be understood to designate scenarios other than DW, which do not depend on a non-vanishing active-sterile mixing to populate the dark matter abundance.}, thereby alleviating the tension with Lyman-$\alpha$ and X-ray measurements. While the idea of freeze-in is quite general and applied in much broader contexts, its implementation in the case of a sterile neutrino inherits the constraints in the neutrino sector and can therefore demonstrate some salient features (Section \ref{sec:sndmfi}). In general, a successful realization of sterile neutrino freeze-in requires a feeble coupling between the sterile neutrino and some particle (generally a scalar beyond the SM) in the thermal bath, with the dark matter abundance gradually built up through this feeble coupling over the lifetime of the Universe. Since both the new particle and the feeble coupling are absent in the sterile neutrino extension of the SM, any framework employing freeze-in of sterile neutrino dark matter is confronted with the task of motivating the existence of such a particle and the appropriately sized feeble coupling. The major purpose of this article is to review such frameworks present in the literature (Section \ref{sec:models}) along with their general characteristics (Section \ref{sec:sndmfi}) and phenomenology (Section \ref{sec:pheno}). For more general aspects of sterile neutrino dark matter, the reader is directed to several excellent reviews in the literature \cite{Boyarsky:2009ix,Abazajian:2012ys, Merle:2013gea,Drewes:2013gca}. 

The article is organized as follows. Section \ref{sec:sndm} discusses sterile neutrino dark matter produced from DW, and constraints from X-ray and Lyman-$\alpha$ measurements. Section \ref{sec:sndmfi} covers general aspects of freeze-in production of sterile neutrino dark matter. Section \ref{sec:models} is devoted to a (non-exhaustive) discussion of a variety of models in the literature that motivate an extended setup where freeze-in of sterile neutrino dark matter can be naturally realized. Section \ref{sec:pheno} examines phenomenological signatures of the freeze-in setup. The main points of the article are summarized in Section \ref{sec:summary}. 


\section{Sterile Neutrino Dark Matter: The Traditional Approach}
\label{sec:sndm}

The (minimal) seesaw mechanism to generate neutrino masses involves extending the SM by right-handed, SM gauge-singlet sterile neutrinos $N_i$, resulting in the following new terms in the Lagrangian
\begin{equation}
\label{eq:seesaw}
\mathcal{L}\supset y_{\alpha i} \bar{L}_\alpha H^\dagger_u N_i+M_i \bar{N}^c_i N_i.
\end{equation}
The first term gives rise to a Dirac mass between the left- and right-handed neutrinos after electroweak symmetry breaking; the second term represents a Majorana mass for the sterile neutrinos (in an appropriately chosen diagonal basis), which is not prohibited by any symmetries if the $N_i$ are complete singlets, as is assumed. If $M\gg y \langle H_u\rangle$, the ensuing seesaw mechanism (known as type-I seesaw) results in active neutrino masses of scale $m_a\approx(y \langle H_u\rangle)^2/M$. 

Two observations are worthy of note here. First, requiring active neutrino masses consistent with oscillation and other astrophysical data ($m_a\sim 0.05$ eV) does not uniquely constrain the Majorana mass scale $M$; GUT scale seesaw models \cite{Minkowski:1977sc, Mohapatra:1980yp, Yanagida:1980xy, GellMann:1980vs, Schechter:1980gr} employ $y\sim\mathcal{O}(1)$ and $M\sim 10^{10}-10^{15}$ GeV, but the Yukawa couplings $y_{\alpha i}$ can be appropriately tuned to give the correct active masses for a wide range of values of $M$. Second, while the most natural choice for the number of sterile neutrinos, in keeping with the number of fermion generations in the SM, is three, only two are required to successful explain the observed neutrino oscillations; this implies that the third sterile neutrino can essentially be decoupled from the seesaw mechanism, which proves to be crucial for its realization as a dark matter candidate. As is the norm, the sterile neutrino dark matter candidate is designated $N_1$, while the remaining two sterile neutrinos responsible for the generation of neutrino masses are labelled  $N_2, N_3$. 

\begin{figure}[t]
\centerline{\includegraphics[width=4.0in]{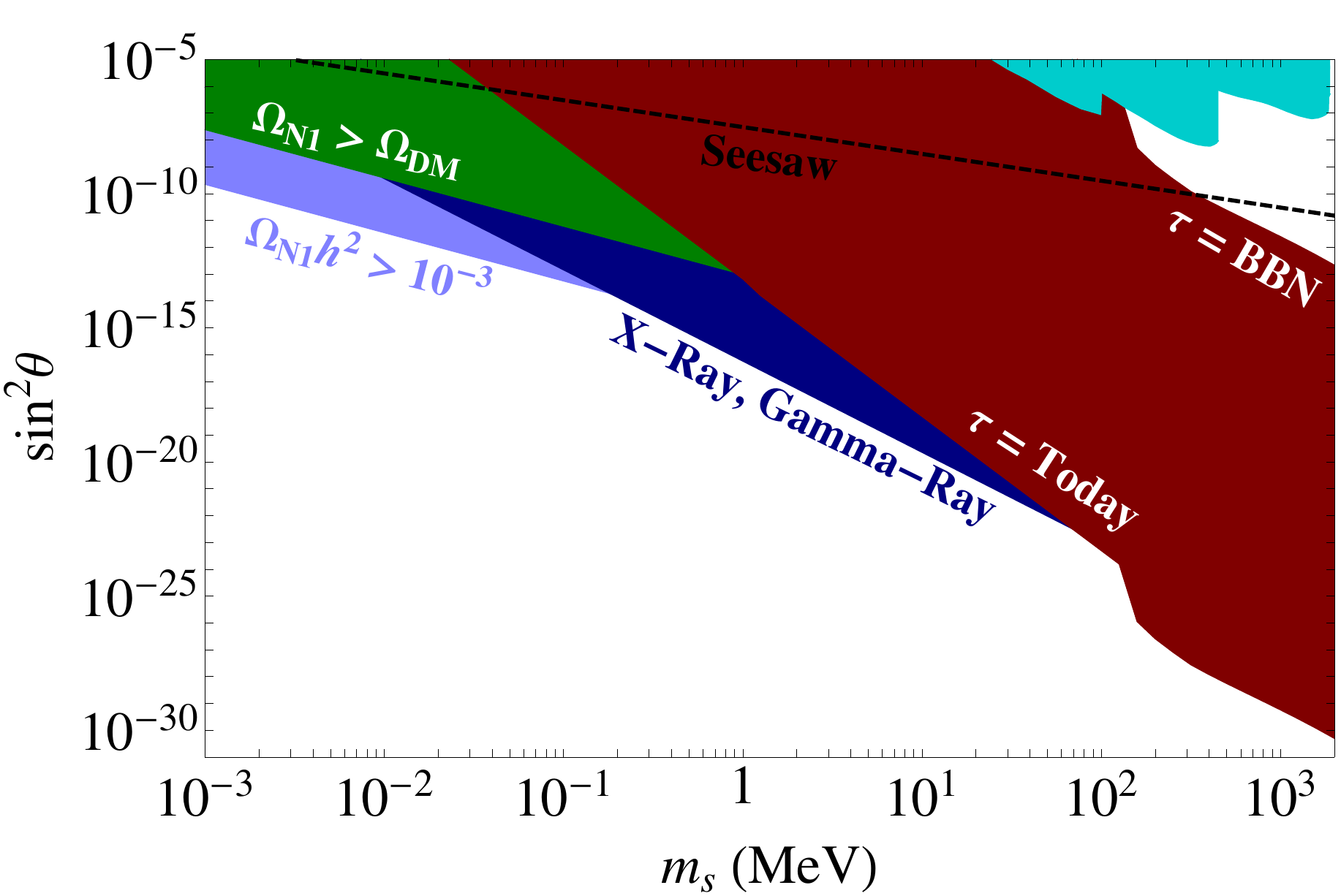}}
\vspace*{8pt}
\caption{Sterile neutrino parameter space. Dotted black: mass-mixing angle combinations enforced by the seesaw mechanism. Light blue: Dodelson-Widrow (DW) mechanism accounts for between 1\% and all of dark matter (DM). Constraints: Red: sterile neutrino lifetime is between BBN ($=1s$) and the current age of the Universe; Green: DW abundance exceeds observed dark matter abundance; Dark blue: X-ray and gamma-ray measurement constraints \cite{Essig:2013goa} for DM abundance as given by DW; Cyan: constraints from direct searches for neutral leptons \cite{PIENU:2011aa,Bergsma:1985is,Ruchayskiy:2011aa,Bernardi:1985ny,Bernardi:1987ek,Vaitaitis:1999wq}. \label{fig1}}
\end{figure}

The mixing between $N_1$ and the active neutrinos generates an $N_1$ abundance through active-sterile oscillation at temperatures $\sim 150$\,MeV; this is known as the Dodelson-Widrow (DW) mechanism \cite{Dodelson:1993je}, or non-resonant production. The relic abundance generated through this mechanism is approximately \cite{Kusenko:2009up,Dodelson:1993je, Abazajian:2005gj,Dolgov:2000ew, Abazajian:2001nj,Asaka:2006nq}
\begin{equation}
\Omega_{N_i} \sim0.2\left(\frac{{\text{sin}}^2 \theta_i}{3\times 10^{-9}}\right)\left(\frac{m_s}{3\,\text{ keV}}\right)^{1.8},
\end{equation}
where sin$^2\theta_i\approx\sum_\alpha y_{\alpha i}^2 \langle H_u\rangle^2/M^2$ is the mixing angle with the active neutrinos. The same mixing can also lead to $N_1$ decay through electroweak processes. The twin requirements of producing the observed dark matter abundance $\Omega_{DM} h^2\!=\!0.12$ from DW and a lifetime longer than the age of the Universe are only satisfied for keV scale or lower masses (see Fig.\,\ref{fig1})\footnote{ For listings of various decay channels and widths of keV-GeV scale sterile neutrinos used to calculate the lifetime in this figure, see e.g. the appendix of Ref.\,\refcite{Gorbunov:2007ak}.}. Note that the required mixing for DM production (light blue region) lies significantly below what would be required for $N_1$ to participate in the seesaw mechanism, denoted by the dotted black line in the figure. Likewise, it is also worth pointing out that the decays of $N_2, N_3$ are constrained by several recombination era observables \cite{Akhmedov:1998qx, Asaka:2005pn, Asaka:2006ek, Kusenko:2009up, Hernandez:2014fha, Vincent:2014rja}, hence they are generally required to decay before Big Bang Nucleosynthesis (BBN), forcing $m_{N_2,N_3}\gtrsim$ a few hundred MeV (see Fig.\,\ref{fig1}). 

This framework of keV scale sterile neutrino dark matter produced through DW and GeV scale $N_2, N_3$ (which can account for baryogenesis) has been extensively studied in  the Neutrino Minimal Standard Model ($\nu$MSM) \cite{Asaka:2005an,Asaka:2005pn,Asaka:2006ek,Canetti:2012vf,Canetti:2012kh}, with this mass pattern possibly emerging from symmetry considerations \cite{Shaposhnikov:2006nn}. A review of the model building aspects of keV sterile neutrino dark matter is presented in Ref. \refcite{Merle:2013gea}.

X-ray measurements provide stringent constraints on sterile neutrino dark matter, since active-sterile mixing causes $N_1$ to decay into an active neutrino and a photon. The decay width for this process is
\begin{equation}
\Gamma(N_1\rightarrow \gamma \nu_a)=\frac{9 \alpha_{EM} G_F^2}{1024\pi^4}\sin^2(2\theta) m_s^5.
\end{equation}
Current X-ray bounds rule out $m_{N_1}\gtrsim 4$ keV \cite{Boyarsky:2008mt,Boyarsky:2006fg,Boyarsky:2006ag, Boyarsky:2005us,Boyarsky:2007ay,Boyarsky:2007ge,Laine:2008pg,Canetti:2012vf,Canetti:2012kh} if the entirely of dark matter consists of sterile neutrinos produced through DW.
 
In the opposite regime, lower mass sterile neutrinos are strongly constrained by considerations of free-streaming length and small scale structure formation. The free-streaming length of sterile neutrino dark matter can be expressed in terms of its mass and momentum \cite{Petraki:2007gq}
\begin{equation}
\Lambda_{FS}\approx 1.2\, \text{Mpc} \left(\frac{\text keV}{m_s}\right) \left(\frac{\langle p_s\rangle}{3.15 T}\right)_{T\approx 1keV}.
\label{fslength}
\end{equation}
As a rough guide, the regimes for cold, warm, and hot dark matter are approximately $\Lambda_{FS}\!\lesssim\!0.01 \text{Mpc}, \,0.01\!\lesssim \!\Lambda_{FS} \!\lesssim \!0.1 \text{Mpc},$ and $ 0.1 \text{Mpc}\!\lesssim\!\Lambda_{FS}$ respectively. An important aspect of keV sterile neutrino DM from DW is that it is warm; for production through DW, its momentum distribution is approximately $(\langle p_s\rangle/3.15 T)_{T\approx 1keV}=0.8-0.9$\cite{Petraki:2007gq}. Compared to WIMP-motivated cold dark matter (CDM) models, a warm candidate offers possible resolution of recent puzzles such as the core vs.\ cusp problem and the ``too big to fail" problem \cite{Lovell:2011rd, BoylanKolchin:2011dk}. However, measurements of the Lyman-$\alpha$ forest, which probes small scale structure formation at various redshifts, have been shown to rule out $m_{N_1}\lesssim 8$ keV \cite{Viel:2006kd, Seljak:2006qw, Asaka:2006nq, Boyarsky:2008xj} if DW-produced sterile neutrino accounts for all of dark matter. A combination of the X-ray and Lyman-$\alpha$ forest data hence rules out the entire window for sterile neutrino dark matter produced through DW (see Ref.\,\refcite{Horiuchi:2013noa} for a recent summary). It should, however, be kept in mind that $N_1$ produced through the DW mechanism can still constitute $\mathcal{O}(10)\%$ of the dark matter abundance \cite{Boyarsky:2008xj,Boyarsky:2008mt}, and even account for the aforementioned 3.5 keV X-ray signal\cite{Harada:2014lma}.

Resurrecting the sterile neutrino as a viable dark matter candidate therefore requires an alternate production mechanism that evades one or both of the X-ray and Lyman-$\alpha$ constraints; several options have been explored in the literature. Possibilities include resonant production \cite{Shi:1998km}, extended frameworks that realize thermal freeze-out followed by entropy dilution \cite{Boyarsky:2008mt,Nemevsek:2012cd,Asaka:2006ek,Asaka:2006nq,Bezrukov:2009th,Patwardhan:2015kga}, interactions of light vector bosons \cite{Shuve:2014doa,Khalil:2008kp}, pion decay \cite{Lello:2014yha}, and neutrino mixing with a hidden dark sector \cite{Falkowski:2011xh}. The remainder of this article will be devoted to a detailed study of another viable alternative: freeze-in production from a feeble coupling to a particle beyond the Standard Model. 


\section{Sterile Neutrino Dark Matter from Freeze-In}
\label{sec:sndmfi}

This section covers general aspects of sterile neutrino dark matter production through the freeze-in mechanism \cite{Chung:1998rq, Hall:2009bx}, commonly referred to as the FIMP framework; specific models will be discussed in the next section. The treatment here contains several approximations for simplicity -- for instance, all analytic formulae presented in this article assume that the number of degrees of freedom during the production of dark matter remains constant; a more careful calculation must take this and several other aspects into account \cite{Merle:2015oja, Drewes:2015eoa}. 

\subsection{Boltzmann Equations}

In the FIMP framework, $N_1$ is assumed to have negligible initial abundance in the early Universe and only a feeble coupling, denoted here by $\lambda$, to the particles present in the thermal bath. This results in a gradual accumulation of $N_1$ as it is slowly produced via this coupling, never attaining equilibrium. The traditional approach to track the abundance and phase space distribution of a species in the early Universe is via the Boltzmann equation
\begin{equation}
\hat{L}[f_X]=C[f_X].
\end{equation}
Here $f_X(p,t)$ is the phase space density of species $X$, $\hat{L}$ is the Liouville operator
\begin{equation}
\hat{L}=\frac{\partial}{\partial t}-H p \frac{\partial}{\partial p},
\end{equation}
with H the Hubble parameter, and the collision term $C[f_X]$ is the source term, incorporating the interactions of $X$ with other particles in the bath. For a generic interaction $a+b+...\leftrightarrow X+A+B+...$, where $a,b,...,A,B,...$ are other species in the thermal bath, the collision term is
\begin{eqnarray}
C[f_X]&=&\frac{1}{2E_X}\int dP_a dP_b...dP_A dP_B...\times (2\pi)^4 \delta^{(4)}(p_a+p_b...-p_X-p_A-p_B...)|M|^2\nonumber\\ 
&& ~~~~\times [f_a f_b... (1\pm f_X)(1\pm f_A) (1\pm f_B)...-f_X f_A f_B...(1\pm f_a)(1\pm f_b)...],
\end{eqnarray}
where $M$ is the matrix element for the interaction, and the $(1\pm f)$ factors correspond to bosonic/fermionic degrees of freedom (this dependence generally turns out to be unimportant and can be dropped). For freeze-in, the $X$ abundance is always sufficiently small that $a+b+...\rightarrow X+A+B+...$ dominates over the inverse process, so the second term in the parenthesis above, proportional to $f_X$, can be dropped. In general, one must solve a coupled system of Boltzmann equations for all species present in the Universe; however, the system simplifies considerably if all but one species are in equilibrium, as is usually the case in scenarios of interest.  

\subsection{(IR) Freeze-In}

The most commonly employed freeze-in production scenario for sterile neutrino dark matter is through the decays of a scalar, $\phi\rightarrow N_1 B$, where $\phi$ represents the scalar and $B$ stands for some other particle. Such decays lead to an accumulation of $N_1$ as long as $\phi$ is abundant in the Universe, approximately until the temperature drops below $\sim m_\phi$. The above Boltzmann equations can be solved for this setup to yield the relic abundance of $N_1$; the resulting abundance is known to be\cite{Hall:2009bx}
\begin{equation}
\Omega_{N_1} h^2\approx \frac{1.09\times 10^{27} g_\phi}{g_*^S \sqrt{g_*^\rho}} \frac{m_{N_1}\Gamma_\phi} {m^2_\phi},
\label{irabundance}
\end{equation}
where $g_\phi,m_\phi,$ and $\Gamma_\phi$ are the internal spin degrees of freedom, mass, and decay width of the scalar $\phi$. $ g_*^S$ and $g_*^\rho$ denote the effective number of degrees of freedom in the bath for the entropy $S$ and energy density $\rho$ respectively; these are to be evaluated at the temperature at which freeze-in occurs (generally $T\sim m_\phi$ in the case of scalar decay). If this occurs at or above the electroweak scale, all SM degrees of freedom are relevant, and $ g_*^S\approx g_*^\rho\approx 100$ is a good approximation unless several degrees of freedom beyond the SM are present. Under this assumption, and writing $\Gamma_\phi\sim\frac{\lambda^2}{8\pi} m_\phi$, the size of the coupling required to obtain the observed dark matter relic density is \cite{Hall:2009bx}
\begin{equation}
\lambda\approx 1.5 \times 10^{-13} \left( \frac{m_\phi}{m_{N_1}}\right)^{1/2}.
\label{couplingsize}
\end{equation}

As discussed in Ref.\,\refcite{Petraki:2007gq}, a useful estimate of the relic abundance is obtained by multiplying the number density of $\phi$ by its decay rate $\Gamma_\phi$ and the time available for this decay to populate $N_1$. Assuming this production channel remains active down to $T\sim m_\phi$ and using the time-temperature relation $t\sim M_0/2T^2$ for a radiation-dominated Universe, where $M_0=\left(\frac{45 M_{Pl}^2}{4\pi^3g_*}\right)^{1/2}\!\!\sim 10^{18}$ is the reduced Planck mass, this estimate yields \cite{Petraki:2007gq}
\begin{equation}
\left(\frac{N_{N_1}}{T^3}\right)|_{T\sim M_\phi}\sim \Gamma_\phi \frac{M_0}{T^2}|_{T\sim M_\phi}\sim \frac{\lambda^2}{8\pi}\frac{M_0}{m_\phi},
\end{equation} 
which is consistent with the expression in Eq.\,\ref{irabundance}. This estimate also makes it clear that the relic abundance of $N_1$ is dominated by physics at the lowest temperatures at which production can occur, i.e. $T\sim m_\phi$. For this reason, this mode of production is also referred to in the literature as IR freeze-in, in contrast to other processes that can be sensitive to UV physics, as will be discussed in the next sub-section.

The above assumes that the $N_1$ abundance accumulates from the decay of $\phi$ while in equilibrium; however, other variations can occur. The scalar can itself freeze-in \cite{McDonald:2001vt,Yaguna:2011qn,Merle:2013wta,Adulpravitchai:2014xna,Kang:2014cia} -- in which case its abundance is also given by a freeze-in calculation -- and eventually decay into $N_1$; likewise, decays to $N_1$ can dominantly occur only after the scalar goes out of equilibrium, in which case the details of the model become important and a generic formula for the eventual $N_1$ abundance cannot be written down. These variants will be discussed in greater detail within a specific framework in Sec.\,\ref{extendedsector}.

Another important aspect of freeze-in is that, compared to DW production, the $N_1$ population is produced at far higher temperatures, much earlier in the cosmological history. This serves an important purpose: dark matter cools down for a longer period, and is insensitive to subsequent entropy injection in the thermal bath, resulting in a cooler spectrum in the present epoch. The momentum distribution for $N_1$ from freeze-in is calculated to be \cite{Kusenko:2006rh, Petraki:2007gq,Petraki:2008ef}
\begin{equation}
\left(\frac{p_s}{3.15\,T}\right)_{T\approx 1keV}\lesssim 0.2
\label{irmomentum}
\end{equation}
for production at or above the weak scale. This is a significantly cooler spectrum than production from DW and therefore more compatible with the Lyman-$\alpha$ bound (see Eq.\,\ref{fslength} and subsequent discussion; also Ref.\,\refcite{Merle:2014xpa}.) \footnote{It is worth keeping in mind, however, that the naive use of average momentum as a measure of compatibility with Lyman-Lyman-$\alpha$ can be misleading if the momentum distribution is nontrivial (see Ref.\,\refcite{Merle:2015oja} for examples and discussions). }.

\subsection{UV Freeze-In}

In models where dark matter interactions involve non-renormalizable operators, freeze-in production can be dominated by interactions at higher temperatures and therefore sensitive to UV physics \cite{Hall:2009bx,Elahi:2014fsa}. This component is generally ignored in most discussions of sterile neutrino dark matter from freeze-in, as only renormalizable interactions are considered. However, given that higher dimensional operators provide a natural framework for realizing extremely small couplings, which are necessary for the realization of (IR) freeze-in, it is always prudent to consider the possibility of a UV freeze-in component (see Ref.\,\refcite{Roland:2014vba},\refcite{Roland:2015yoa} for an illuminating example). 

For concreteness, the following dimension five operator involving a sterile neutrino
\begin{equation}
\mathcal{L}\supset \frac{\lambda}{M}\phi_1\phi_2 N_1 B,
\end{equation}
where $\phi_1,\phi_2$ are scalars, $B$ is a fermion, and $M$ is the cutoff scale of the theory, results in the production of $N_1$ through interactions such as $\phi_1 \phi_2\rightarrow N_1 B$. The dark matter yield from this interaction is \cite{Hall:2009bx}
\begin{equation}
Y_{UV}\approx\frac{0.4\, T_R \,\lambda^2 M_{Pl}}{\pi^7 M^2\, g_*^S \sqrt{g_*^\rho}}
\end{equation}
where $T_R$ is the reheat temperature. Assuming, again, that all production occurs above the electroweak scale, the $N_1$ relic density can be written as \cite{Hall:2009bx,Elahi:2014fsa}
\begin{equation}
\Omega_{N_1} h^2\sim 0.1\,\lambda^2\left(\frac{m_{N_1}}{\rm{MeV}}\right)\left(\frac{T_{R}\,M_{Pl}}{M^2}\right).
\end{equation}
Compared to the analogous expression from IR freeze-in, Eq.\,\ref{irabundance}, there are three key differences: 
\begin{itemlist}
 \item It has explicit dependence on $T_{R}$: it is sensitive to physics at the earliest temperatures.
 \item It is independent of $m_\phi$: production is dominant at high temperatures, where $\phi$ is relativistic, and turns off at low temperatures, so that the point where it terminates -- represented by $T\sim m_\phi\sim m_{\phi_1},m_{\phi_2}$ -- is irrelevant.
 \item The coupling $\lambda$ can be $\mathcal{O}(1)$, depending on the values of $T_R$ and $M$; this is in stark contrast to the IR case, where $\lambda$ is required to be extremely feeble (Eq.\,\ref{couplingsize}). 
\end{itemlist} 
On the other hand, the average momentum for $N_1$ produced via UV freeze-in is expected to be comparable to that from IR freeze-in. While this population is produced at much higher temperatures, it redshifts along with the rest of the components in the radiation-dominated Universe, therefore maintaining the same temperature until $T\sim m_\phi$, at which point the subsequent evolution maps onto that from IR freeze-in. Eq.\,\ref{irmomentum} is therefore also representative of UV freeze-in.

\subsection{Entropy Dilution}

An important aspect of calculating the abundance and momentum distribution of sterile neutrino dark matter is taking into account any entropy that is injected into the thermal bath between the production of the dark matter population and the present epoch. Since $N_1$ is out of equilibrium from the moment of production, such entropy injection produces the effect of diluting the $N_1$ abundance and redshifting it relative to the visible sector. 

In particular, the decoupling of the SM degrees of freedom results in the dilution of the $N_1$ relic density by a factor of $\xi$ and an additional redshift of the momentum $\langle p_s \rangle$ by a factor $\xi^{1/3}$ \cite {Kusenko:2006rh}, where $\xi=g_*(m_\phi)/g_*(\rm keV)$ is the reduction in the number of degrees of freedom from the time of production $T\sim m_\phi$ to the present epoch; if production occurs above the electroweak scale, $\xi\approx100/3\approx33$. This dilution has been accounted for in the formulae in the previous subsections. A first order phase transition after dark matter production can also dilute and redshift the $N_1$ population; this is studied in detail in Ref.\,\refcite{Petraki:2007gq}. Likewise, any heavy, long-lived BSM particle that decays after DM production can cause non-negligible entropy dilution. This is of particular interest for sterile neutrino dark matter since the minimal sterile neutrino extension of the SM already contains two BSM particles, the heavier sterile neutrinos $N_{2,3}$; if their masses are not too far above $N_1$, as would be plausible if all $N_i$ Majorana masses share a common origin, the above criteria are satisfied, and the entropy release from their decays deserves special attention.

The ratio of entropy from $N_{2,3}$ decays to the entropy in the remainder of the system, which provides the suppression factor for the dark matter relic density, is calculated to be  \cite{Scherrer:1984fd,Bezrukov:2009th,Asaka:2006ek}
\begin{equation}
S\approx \left(1+2.95\left(\frac{2\pi^2\bar{g}_*}{45}\right)^{1/3}\left(\frac{r^2 M_N^2}{M_{Pl} \Gamma_N},\right)^{2/3}\right)^{3/4}
\label{sns}
\end{equation}
where $m_N$ and $\Gamma_N$ are the mass scale and decay width of the sterile neutrino $N$ ($N$ stands for $N_2$ or $N_3$), $\bar{g}_*$ is the average effective number of degrees of freedom during $N$ decay, and $r$ is the $N$ abundance when it decouples, given by \cite{Scherrer:1984fd,Bezrukov:2009th}
\begin{equation}
r\equiv \frac{n_N}{s}=\frac{135\, \zeta(3)}{4\pi^4g_*},
\end{equation}
where $g_*$ represents the number of degrees of freedom when $N$ freezes out. If $S$ is sufficiently larger than $1$, the first term in Eq.\,\ref{sns} can be dropped to yield a simpler expression \cite{Bezrukov:2009th}
\begin{equation}
S\approx 0.76 \frac{\bar{g}_*^{1/4}M_N}{g_*\sqrt{\Gamma_N\,M_{Pl}}}
\end{equation}
The numerical value of $S$ can be estimated by approximating $\Gamma_N\sim G_F^2 \,M_N^5\,\theta_N^2/(192\pi^3)$ \cite{Bezrukov:2009th}, with $\theta$ as specified by the requirement of the seesaw (see Fig.\,\ref{fig1}), and using the information that $N_{2,3}$ decouple around $\mathcal{O}(20)$ GeV \cite{Asaka:2006ek}; this yields $S\sim\mathcal{O}(1)$ for $N_{2,3}$ at the GeV scale. 

\subsection{Characteristic Features of Freeze-In}

Before delving into details of specific models, it is worth highlighting the following features that are characteristic of sterile neutrino dark matter from freeze-in and in stark contrast to DW production. 
\begin{itemlist}
\item Since the production mechanism does not rely on active-sterile mixing, the mixing angle $\theta_1$ between $N_1$ and the active neutrinos can be arbitrarily small, and even vanish. A vanishing $\theta_1$ is technically natural, since it leads to a $\mathcal{Z}_2$ symmetry for $N_1$. In this limit, $N_1$ is perfectly stable, eliminating the prospects of observing any astrophysical signal from its decay \cite{Boyarsky:2012rt}. A non-vanishing mixing angle is constrained by X-ray and gamma-ray data as well as the requirement that the $N_1$ lifetime exceed the age of the Universe (see Fig.\,\ref{fig1}).
\item A testable prediction of a vanishing $\theta_1$ is that the lightest active neutrino is essentially massless; two sterile neutrinos in the seesaw can only give masses to two active neutrinos, leaving the third one massless (this is dubbed the seesaw fair play rule\cite{Xing:2007uq}). The lightest neutrino does, however, receive a mass at 2 loops from diagrams involving SM fields \cite{Davidson:2006tg}. 
\item Another consequence of the relaxation on the requirement of $\theta_i$ is that $N_1$ is no longer constrained to be at the keV scale. Any mass that satisfies the relevant constraints is acceptable, significantly broadening the allowed window for sterile neutrino dark matter. 
\item It is worth reiterating that $N_1$ from freeze-in is generally produced at earlier temperatures compared to DW, and therefore tends to have a colder spectrum. While sterile neutrinos are traditionally thought of as warm dark matter, depending on details of the freeze-in process and the energy scales of the particles involved, this mechanism can result in cold dark matter. 
\end{itemlist}


\section{Models for Freeze-In of Sterile Neutrino Dark Matter}
\label{sec:models}

While the previous section painted the general framework of sterile neutrino dark matter from freeze-in, explicit models are necessary to complete the story, as the setup requires an extension beyond the minimal $\nu$MSM content. A successful realization of $N_1$ freeze-in requires a feeble coupling to a BSM particle present in the early Universe, neither of which are a-priori present in the neutrino sector. Such content has been motivated from various considerations in the literature. This section is devoted to a study of a variety of such scenarios where freeze-in of sterile neutrino dark matter can naturally occur. It must be emphasized that this is not intended to be an exhaustive list of all possibilities discussed in the literature, but only a meaningful sample of the variety of ways the desired framework can emerge. 

\subsection{Coupling to the Inflaton}
Given the plethora of evidence supporting an inflationary phase in the earliest moments of the Universe, the inflaton is a well-motivated BSM scalar. A framework extending the $\nu$MSM by a single real scalar field to account for inflation, the inflaton, whose decay also produces a relic abundance of sterile neutrino dark matter, was presented in Ref.\,\refcite{Shaposhnikov:2006xi}. The model employs the following scale-invariant Lagrangian:
\begin{equation}
\mathcal{L}=\mathcal{L}_{\nu MSM [M\rightarrow 0]}+\frac{1}{2}(\partial_\mu \chi)^2-\frac{f_I}{2}\bar{N}_I^c N_I \chi+h.c.-V(\Phi,\chi)
\end{equation}
Here $\chi$ is the inflaton, $\Phi$ represents the Higgs doublet, and the sterile neutrinos $N_I$ have Yukawa interactions with the inflaton with strength $f_I$. The scalar potential is assumed to contain the most general scale-invariant terms, augmented by an inflaton mass term (necessary for electroweak symmetry breaking \cite{Buchmuller:1990pz}) that explicitly breaks the scale invariance
\begin{equation}
V(\Phi,\chi)=-\frac{1}{2}m_\chi^2 \chi^2+\lambda(\Phi^\dagger \Phi-\frac{\alpha}{\lambda}\chi^2)^2+\frac{\beta}{4}\chi^4
\end{equation}
The resulting inflaton vacuum expectation value (vev) and mass are
\begin{equation}
\langle\chi\rangle\sim m_H/2\sqrt{\alpha},~~~m_I\sim m_H\sqrt{\beta/2\alpha}.
\end{equation}

Depending on the exact values of the parameters (see also Ref.\,\refcite{Anisimov:2008qs} for constraints on parameters), the inflaton mass can be comparable to or lighter than the Higgs, and the inflaton vev is $10^5-10^8$ GeV \cite{Shaposhnikov:2006xi}. The inflaton can remain in thermal equilibrium down to $T\,\textless\, m_I$ thanks to interactions induced by Higgs-inflaton mixing, hence its decays can produce the correct abundance of sterile neutrinos to explain dark matter with an appropriate choice of the Yukawa coupling $f_1$. Note that the Yukawa term is also responsible for giving rise to the sterile neutrino masses. For instance, $m_I\sim 300\,$MeV ($100\,$GeV) implies $f_1\approx 10^{-10}$ and $m_{N_1}\sim 20\,$keV ($\,\mathcal{O}(10)\,$MeV) to produce the required dark matter abundance\cite{Shaposhnikov:2006xi}. A similar setup was also studied in Ref.\,\refcite{Bezrukov:2009yw}, deriving the dark matter abundance
\begin{equation}
\Omega_N=\frac{1.6 f(m_\chi)}{S}\frac{\beta}{1.5\times 10^{-13}}\left(\frac{M_{N_1}}{10\,\rm{keV}}\right)^3\left(\frac{100\,\rm{MeV}}{m_\chi}\right),
\end{equation}
where $S$ accounts for entropy dilution from decays of heavier sterile neutrinos, and $f(m_\chi)$ is determined by the number of degrees of freedom during inflaton decay. 

In such frameworks, the requirement that the flatness of the inflaton potential not be spoiled by radiative corrections constrains $f_I\lesssim 2\times 10^{-3}$; however, producing the correct dark matter abundance through freeze in requires much smaller couplings $\mathcal{O}(10^{-10})$.

Along a different direction, production of sterile neutrino dark matter from decays of the inflaton-Higgs condensate during preheating was studied in Ref.\,\refcite{Bezrukov:2008ut}.

\subsection{Extended Higgs Sector}
\label{extendedsector}
Several papers in the literature have studied $N_1$ freeze-in from an extended Higgs sector, motivated by the possibility that the sterile neutrino Majorana masses arise from the Higgs mechanism, analogous to the SM fermion masses. This is reminiscent of the coupling to the inflaton discussed in the earlier subsection, but without the additional constraints from inflation. These models extend the Higgs sector with a SM singlet Higgs boson $S$, which can mix with the SM Higgs $H$. This subsection will be based on the discussions in Ref.\,\refcite{Petraki:2007gq},\,\refcite{Kusenko:2006rh}, and \refcite{Merle:2013wta}; however, freeze-in of sterile neutrino dark matter with similar extensions of the Higgs sector has also been explored in, e.g. a neutrophilic two Higgs doublet model \cite{Adulpravitchai:2015mna}, the scotogenic model \cite{Molinaro:2014lfa}, a scale invariant extension of the $\nu$MSM where the scalar also helps with EWSB\cite{Kang:2014cia}, and a general approach with the aim of improving leptogenesis scenarios in addition to improving the dark sector \cite{Matsui:2015maa}. Possible symmetries behind the required mass and Yukawa coupling patterns in such setups were explored in Ref.\,\refcite{Allison:2012qn}.

In general, the extension of the Higgs sector by the scalar $S$ introduces the following additional terms in the Lagrangian \cite{Petraki:2007gq}:
\begin{equation}
\mathcal{L}\supset -\frac{h_a}{2}S \bar{N}_a^c N_a+V(H,S)+h.c.
\label{higgsextension}
\end{equation}
The first terms leads to Majorana masses for the sterile neutrinos $N_a$ once $S$ obtains a vev. The scalar potential is \cite{Petraki:2007gq}
\begin{equation}
V(H,S)=\mu_H^2|H|^2+\frac{1}{2}\mu_S^2 S^2-\frac{1}{6}\alpha S^3-\omega |H|^2 S-2\lambda_{HS} |H|^2 S^2-\frac{1}{4}\lambda_S S^4-\lambda_H |H|^4.
\end{equation}
Ref.\,\refcite{Kusenko:2006rh,Merle:2013wta,Merle:2015oja} impose an additional symmetry that eliminates the cubic terms, i.e. $\alpha=\omega=0$, leading to a simplified framework where the $|H|^2 S^2$ term acts as the sole portal between the Higgs bosons. After electroweak symmetry breaking, both $H$ and $S$ obtain weak scale masses and vevs. In this model, the electroweak phase transition can be first-order, resulting in entropy production that can suppress this abundance by at most a factor of 1.3\cite{Kusenko:2006rh,Enqvist:1992va}. If $h_a$ (in Eq.\,\ref{higgsextension}) for the lightest sterile neutrino $N_1$ is sufficiently small, freeze-in production of $N_1$ occurs through the decay $S\rightarrow N_1 N_1$; the production of $S$ and subsequent freeze-in of $N_1$, however, can occur in several ways.

\subsubsection{Decay of Scalar in Equilibrium}

Interactions in the Higgs sector \cite{McDonald:1993ex}, if sufficiently strong, can keep $S$ in equilibrium with the thermal bath: equilibrium is attained if $\lambda_{HS}\gtrsim 10^{-6}$ \cite{Kusenko:2006rh,Petraki:2007gq}, or $\alpha\,\omega/m_S^2\gtrsim \lambda_{HS}$ if the more general form of the potential with cubic terms is retained\cite{Petraki:2007gq}. Sterile neutrino production from in (or out of) equilibrium decays of $S$ has been studied in detail in Ref.\,\refcite{Kusenko:2006rh},\refcite{Petraki:2007gq}. If most of the $N_1$ abundance is produced at temperatures of order $m_S\sim (0.1-1)$\,TeV while $S$ is still in equilibrium, the ensuing relic density is calculated to be\cite{Kusenko:2006rh,Petraki:2007gq}
\begin{equation}
\Omega_{N_1} \sim 0.2 \left(\frac{h_1}{1.4\times10^{-8}}\right)^3 \left(\frac{\langle S \rangle}{m_S}\right)\sim 0.2 \left(\frac{h_1}{1.4\times10^{-8}}\right)^3 \left(\frac{1}{\sqrt{2\lambda_S}}\right).
\end{equation}
Ref.\,\refcite{Kusenko:2006rh},\refcite{Petraki:2007gq} do not provide any explanation for the smallness of $h_1\sim 10^{-8}$ required to produce the correct relic density.

For the sake of comparison with the subsections to follow, it is also useful to look at the $N_1$ yield \cite{Merle:2015oja}
\begin{equation}
Y_{N_1}(r\rightarrow\infty)\sim\frac{135\, h_1^2}{1024\,\pi^3\, g_*(T_{\rm prod})}\frac{M_0}{m_S}.
\end{equation}
Recall that the relic abundance can be calculated from this as
\begin{equation}
\Omega h^2=\frac{m_{N_1} Y_{N_1}(r\rightarrow\infty) s_0}{\rho_c/h^2},
\end{equation}
where $s_0\approx2900$cm$^{-1}$ is the present entropy density, and the critical energy density $\rho_c/h^2\approx10^{-2}$\,MeV\,cm$^{-3}$.

The average momentum in this case is calculated to be\cite{Petraki:2007gq}
\begin{equation}
\left(\frac{\langle p \rangle}{T}\right)_{T\ll \rm{MeV}}\approx 0.75 
\end{equation}

\subsubsection{Out of Equilibrium Decay}

The scalar $S$ can also be sufficiently weakly coupled that it goes out of equilibrium, and the primary $N_1$ production occurs at lower temperatures $T\,\textless\, 100$\,GeV; this occurs for $\alpha,\omega\approx 0,\,\lambda_{H,S}\approx 10^{-6}$. With $\lambda_{H,S}\approx 10^{-6}$ and $m_{N_1}\sim$keV, sufficient amounts of dark matter can be produced \cite{Kusenko:2006rh, Petraki:2007gq}; the $N_1$ yield from such out of equilibrium decays of the scalar is\cite{Merle:2015oja}
\begin{equation}
Y_{N_1}(r\rightarrow\infty)\sim\frac{45}{4\pi^4}\frac{r^2_{FO} K_2(r_{FO})}{g_*(T_{\rm prod})}, 
\end{equation}
Here $K_2$ is the modified Bessel function of the second kind, and $r_{FO}=m_S/T_{FO}$ introduces dependence on the temperature at which the scalar freezes out. The slightly delayed $N_1$ production compared to in-equilibrium production can result in a warmer spectrum for the sterile neutrino population; for instance, for decoupling around $r_{FO}\sim 1-2$, the average momentum was found to be \cite{Petraki:2007gq}
\begin{equation}
\left(\frac{\langle p \rangle}{T}\right)_{T\ll \rm{MeV}}\approx 0.8 
\end{equation}
Results from variations across a wider range of parameters are presented in Ref.\,\refcite{Petraki:2007gq}.

\subsubsection{Freeze-In of Scalar}
If the scalar interactions with the SM are sufficiently feeble that equilibrium is never obtained, $\lambda_{H,S}\ll 10^{-6}$, the scalar itself freezes-in  \cite{McDonald:2001vt,Yaguna:2011qn,Merle:2013wta,Adulpravitchai:2014xna,Kang:2014cia}. The production of sterile neutrino dark matter from subsequent decays of such a frozen-in scalar has been studied in detail in Ref.\,\refcite{Merle:2013wta}. The dark matter yield from such a configuration is approximately\cite{Merle:2015oja}
\begin{equation}
Y_{N_1}(r\rightarrow\infty)\sim\frac{135\, \lambda_{H,S}^2}{1024\,\pi^5\, g_*(T_{\rm prod})}\frac{M_0}{m_S}.
\end{equation}
The typical dark matter particle momentum $\langle p \rangle$ in this case depends on how rapidly the population of $S$ is converted into sterile neutrinos; if this conversion occurs long after the scalar becomes non-relativistic, the final $N_1$ population can be significantly warmer than that produced from the scenarios above. A comprehensive numerical treatment is performed in Ref.\,\refcite{Merle:2015oja}, where all possibilities -- hot, warm, or cold dark matter -- were encountered in various parts of the parameter space. 

\subsection{Non-singlet Sterile Neutrinos, and Supersymmetry}
Ref.\,\refcite{Roland:2014vba},\,\refcite{Roland:2015yoa} considered a supersymmetric framework where the SM-singlet sterile neutrinos $N_i$ are charged under some new symmetry of nature, such as a $U(1)'$. This symmetry prohibits the terms in Eq.\,\ref{eq:seesaw}, eliminating the possibility of obtaining neutrino masses from the traditional seesaw mechanism. Higher dimensional operators involving the SM and $N_i$ fields can be obtained by coupling the $N_i$ to other fields charged under the $U(1)'$; a new field $\phi$ that carries the opposite charge under $U(1)'$ is introduced for this purpose. With chiral supermultiplets $\mathcal{N}_i$ for the sterile neutrinos and a chiral supermultiplet $\Phi$, whose spin $(0,~1/2)$ components are labelled $(\tilde{N}_i,N_i)$ and $(\phi, \psi_\phi)$ respectively, the superpotential contains the following higher dimensional operators:
\begin{equation}
\label{eq:newterms}
W\supset\frac{y}{M_*} L H_u \mathcal{N}\Phi+\frac{x}{M_*}\mathcal{N}\mathcal{N}\Phi\Phi .
\end{equation}
Here $x$ and $y$ are dimensionless couplings (flavor indices suppressed), and $M_*$ is the scale for UV completion, such as $M_{GUT}$ or $M_{Pl}$. If the scalar $\phi$ obtains a vev, this breaks the $U(1)'$, recovering an effective $\nu$MSM like framework with the following active-sterile Dirac mass and sterile Majorana mass terms after electroweak symmetry breaking:
\begin{equation}
m_D=\frac{y \pv\hv}{M_*},~~~~~m_M=\frac{x \pv^2}{M_*}.
\end{equation}
Assuming $\pv\gg \hv$, the seesaw mechanism gives the following sterile and active neutrino masses:
\begin{equation}
\label{Mas}
m_s =m_M=\frac{x \pv^2}{M_*},~~~m_a = \frac{m_D^2}{m_M}=\frac{y^2 \hv^2}{x M_*}.
\end{equation}
This framework therefore involves a modified Higgs mechanism to give mass to the sterile neutrinos compared to the scenarios discussed in the previous subsections.

For $\mathcal{O}(1)$ couplings and $M_*=M_{GUT}$, keV-GeV scale sterile neutrinos are obtained for $\pv\sim 1-100$\,PeV. This framework therefore requires new physics at the PeV scale. The leading candidate for BSM physics is supersymmetry (SUSY), and while a natural resolution of the hierarchy problem dictates that SUSY must be at the weak scale, this is now is strong tension with several null searches; meanwhile, intermediate (PeV) scale SUSY is favored by flavor, CP, and unification considerations \cite{Wells:2003tf, ArkaniHamed:2004fb, Giudice:2004tc, Wells:2004di}, as well as being consistent with a 125 GeV Higgs \cite{Giudice:2011cg, ArkaniHamed:2012gw, Arvanitaki:2012ps}. The vev of $\phi$ and the breaking of $U(1)'$ can therefore plausibly be tied to the scale of supersymmetry breaking.

This framework allows for both IR and UV production of $N_1$ if $\phi$ has additional interactions (with the supersymmetric sector) that keep it in equilibrium in the early Universe. IR production occurs through the decay $\phi\rightarrow N_1\,N_1$ with an effective coupling $x_{1}=\frac{2\, x\, \pv}{M_*}$ once $\phi$ obtains a vev; the relic abundance is \cite{Roland:2015yoa,Roland:2014vba} 
\begin{equation}
\Omega_{N_1} h^2\sim 0.1 \left(\frac{x_1}{1.4\times10^{-8}}\right)^3 \left(\frac{\pv}{m_{\phi}}\right).
\end{equation}
Note that the size of the feeble coupling $x_1\sim 10^{-8}$ required for freeze-in production and the correct relic density arises naturally in this model as an effective coupling generated from a higher dimensional operator, $x_1\sim \pv/M_*$. Likewise, the terms in Eq.\,\ref{eq:newterms} lead to UV production of $N_1$ through the annihilation processes $\phi\,\phi\rightarrow N_1\,N_1$, $\phi\,H_u\rightarrow \nu_a\,N_1$, $\phi\,\nu_a\rightarrow H_u\,N_1$, and $H_u\,\nu_a\rightarrow \phi\,N_1$ (a similar process $HH\rightarrow N N$ from higher dimensional operators to produce sterile neutrino dark matter after reheating (unrelated to the seesaw) was studied in Ref.\,\refcite{Bezrukov:2008ut}). The contribution from $\phi\,\phi\rightarrow N_1\,N_1$ to the dark matter relic density, for instance, is approximately \cite{Roland:2015yoa,Roland:2014vba} 
\begin{equation}
\Omega_{N_1} h^2\sim 0.1\,x^2\left(\frac{m_s}{\rm{GeV}}\right)\left(\frac{1000\,T_{R}\,M_{Pl}}{M_*^2}\right).
\end{equation}
As pointed out in the previous section, this abundance is sensitive to the reheat temperature $T_{R}$.  Finally, it should be noted that the annihilation process $H_u\,\nu_a\rightarrow \phi\,N_{2,3}$ can build up a freeze-in abundance of $\phi$ even if it is not present in the early thermal bath, and its decays can populate $N_1$; the relic abundance from this process is
\begin{equation}
\Omega_{N_1} h^2\sim 0.1\sum_{i,j}\,y_{ij}^2\left(\frac{m_s}{\rm{GeV}}\right)\left(\frac{1000\,T_{R}\,M_{Pl}} {M_*^2}\right)\, Br(\phi\rightarrow N_1 N_1)
\end{equation}

\subsection{Leptogenesis and Freeze-In via a Charged Scalar}
\label{chargedscalar}
An interesting aspect of sterile neutrinos is the possibility of leptogenesis; this can be triggered either by sterile neutrino decay if they lie above the TeV scale \cite{Fukugita:1986hr}, or sterile neutrino oscillations if their masses lie at the GeV scale\cite{Akhmedov:1998qx, Asaka:2005pn}. The latter requires a strong mass degeneracy between $N_2$ and $N_3$, and therefore a high degree of tuning \cite{Canetti:2012kh,Shuve:2014zua}. It has been shown that the presence of a charged scalar $\delta^+$ -- naturally realized in left-right symmetric models and unified extensions of the SM -- with $N_i$ at a few TeV and heavier than $\delta^+$ can help to generate the observed baryon asymmetry via leptogenesis \cite{Frigerio:2006gx} \footnote{For a different variation involving leptogenesis and a neutral scalar, see Ref.\,\refcite{Kang:2014mea}.}.

Ref.\,\refcite{Frigerio:2014ifa} studied freeze-in production of $N_1$ in such a framework, where an $SU(2)$ singlet $\delta^+$ with unit electric charge has the following interactions:
\begin{equation}
\mathcal{L}_\delta \supset - \lambda_{\delta H} \delta^+\delta^- H^\dagger H - \overline{l_{L\alpha}} (y_L)_{\alpha\beta} (i\sigma_2) (l_{L\beta})^c \delta^+ -\overline{(e_{R\alpha})^c} (y_R)_{\alpha i} N_{Ri} \delta^+ ~.
\label{Ldelta}\end{equation}
Generally, leptogenesis from N-decay requires an out of equilibrium decay $N_i\rightarrow H l_{L\alpha}$, which generates a lepton asymmetry; in the presence of $\delta^+$, the analogous decay $N_i\rightarrow \delta^+ e_{R\alpha}$ can also source leptogenesis. Likewise, leptogenesis from N-oscillation involves coherent oscillations of $N_i$ violating lepton flavor numbers and possibly CP-symmetry; this asymmetry is then transferred to $l_L$ provided the interaction $ \overline{(l_{L\alpha})}(y_\nu)_{\alpha i} N_{Ri} H$ equilibrates. In the presence of $\delta^+$, the asymmetries can be transferred to the SM lepton singlets $e_{R\alpha}$ via the $y_R$ coupling. Having both the left- and right- handed leptons participate through the couplings $y_\nu$ and $y_R$ allows greater freedom to generate larger asymmetries; this is particularly true since the former is constrained by the seesaw mechanism, while the latter is not \cite{Frigerio:2014ifa}.

For $m_{N_1}\ll m_{\delta^+}$, the decays of  $\delta^+$ can lead to freeze-in production of $N_1$. Note that $\delta^+$ is in thermal equilibrium due to its interactions with the Higgs and leptons, and produces $N_1$ via $\delta^+\rightarrow N_1 l^+$, resulting in a relic density \cite{Frigerio:2014ifa}
\begin{equation}
\Omega_{N_1} h^2\approx 0.11\left(\frac{m_{N_1}}{\rm keV}\right)\left(\frac{y_{R_1}}{5\times 10^{-8}}\right)^2\left(\frac{\rm TeV}{{M_\delta^+}}\right).
\end{equation}
This estimate was found to be in agreement with numerical calculations\cite{Klasen:2013ypa, Molinaro:2014lfa}. In such models, $N_1$ is a good dark matter candidate only up to $m_{N_1}=1$\,MeV; for heavier masses, it can decay through $N_1\rightarrow \nu e^+ e^-$ via the $y_L$ and $y_R$ couplings even for zero active-sterile mixing and is therefore not sufficiently long-lived. 

\subsection{Warped Extra Dimensions}
Warped extra dimensions represent a well-motivated theory of BSM physics. The radion, a scalar that features in models with extra dimensions and couples to all degrees of freedom in the theory, can play the role of the scalar whose decays populate sterile neutrino dark matter; this possibility was explored in Ref.\,\refcite{Kadota:2007mv} (for a different extra-dimension realization of sterile neutrino dark matter where production occurs via a $B-L$ gauge boson, see Ref.\,\refcite{Kusenko:2010ik}). The radion framework consists of the radion localized on the IR(TeV) brane and Majorana mass terms for sterile neutrinos confined to the UV(Planck) brane; in the 4D effective theory, they interact via a dimensionless effective coupling $\lambda$ (the reader is referred to Ref.\,\refcite{Kadota:2007mv} for details). Gauge interactions thermalize the radion in the early Universe, and if $\lambda$ is not too large, freeze-in production of the dark matter candidate $N_1$ occurs via decays of the radion. The correct relic density is calculated to be obtained for \cite{Kadota:2007mv}
\begin{equation}
\lambda^2\sim 0.3\times 10^{-20} \left(\frac{\rm MeV}{m_{N_1}}\right)\left(\frac{m_r}{100\,\rm{GeV}}\right)
\end{equation}
This is an extremely small coupling, as is characteristic of the freeze-in mechanism. Extra dimension models, however, provide a natural setting to realize such small couplings through suppressed wavefunction overlap in the extra dimension. For some examples of parameter choices where the desired masses and couplings necessary for $N_1$ freeze-in are realized, the reader is referred to Ref.\,\refcite{Kadota:2007mv}.

\subsection{Pseudo-Dirac Neutrinos}
Dark matter can also freeze-in from decays of pseudo-Dirac neutrinos instead of a scalar. This possibility is discussed in Ref.\,\refcite{Abada:2014zra} in the context of the 2,3 inverse seesaw (ISS) framework, which is different from the type I seesaw framework that this article has focused on. In particular, this setup consists of pseudo-Dirac neutrinos $N_I$ that are abundant in the early Universe due to their efficient Yukawa interactions. If they are heavier than the Higgs, their decay can lead to freeze-in of DM via $N_I\rightarrow h+DM$ with an effective coupling $Y_{\rm eff} \sin\theta$; as a Yukawa coupling suppressed by the active-sterile mixing angle $\theta$, this effective coupling can naturally be very small, as required for freeze-in. The DM relic density can be expressed as \cite{Abada:2014zra}
\begin{equation}
\Omega h^2\approx 0.2\left(\frac{\sin \theta}{10^{-6}}\right)^2 \left(\frac{m_{N_1}}{\rm keV}\right)\sum_I g_I \left(\frac{Y_{\rm{eff,I}}}{0.1}\right)^2 \left(\frac{\rm TeV}{m_I}\right) \left(1-\frac{m_h^2}{m_I^2}\right) \epsilon(m_I).
\end{equation}
The function $\epsilon(m_I)$ and coupling $Y_{eft}$ are defined in Ref.\,\refcite{Abada:2014zra}; taking some other considerations into account, the appropriate dark matter relic density was found to be attainable for $m_h\,\textless\, m_I\,\textless\, 1.4$\,TeV. 


\section{Phenomenology}
\label{sec:pheno}

Following discussions of the generic framework and specific realizations of sterile-neutrino dark matter freeze-in in the previous sections, this section discusses various observable aspects of this mechanism. It is important to keep in mind that the phenomenological aspects of sterile neutrino dark matter from freeze-in are significantly different from those from DW. In general, $N_1$ interacts very weakly and is difficult to probe experimentally; nevertheless, some observable aspects do exist. 

\textit{Indirect Detection}: \\ 
Traditionally, the detection of an X-ray line signal at half the sterile neutrino mass from $N_1\rightarrow \nu_a\,\gamma$ is heralded as the ``smoking-gun" signal of sterile neutrino dark matter\cite{Boyarsky:2012rt}. This statement, however, is only applicable for DW production, where a large active-sterile mixing angle is necessary for dark matter production. For production through freeze-in, the mixing angle can be arbitrarily small, and even vanish, hence the detection of such a signal, while possible, is no longer motivated. Indeed, freeze-in can be realized without any direct coupling between $N_1$ and SM, and $N_1$ can be completely stable, in which case there are no dark matter decay or annihilation signals. 

\textit{Direct Searches}:\\
While there is very little hope of directly producing $N_1$ in laboratory experiments because of its feeble coupling to the SM, efforts to directly produce and detect the heavier sterile neutrinos $N_{2,3}$ \cite{PIENU:2011aa,Bergsma:1985is,Ruchayskiy:2011aa,Bernardi:1985ny,Bernardi:1987ek,Vaitaitis:1999wq} can plausibly probe their existence, although this would only amount to a confirmation of the existence of a neutral lepton, not of freeze-in production of sterile neutrino dark matter.

\textit{Active Neutrino with Vanishing Mass}: \\
As mentioned earlier, a direct consequence of the longevity and feeble coupling of $N_1$ is that the lightest active neutrino must be essentially massless (unless there are more than three sterile neutrinos). Observational confirmation of an essentially vanishing active neutrino mass, which is a necessary but not sufficient condition, would therefore lend some credence to the frozen-in sterile neutrino dark matter paradigm. 

\textit{Collider Signals}: \\
The heavy BSM particle that causes $N_1$ to freeze-in, on the other hand, might be accessible to collider probes if it is not too heavy. The charged scalar discussed in Sec.\,\ref{chargedscalar}, for instance, can be pair-produced via Drell-Yan with an off-shell $\gamma/Z$, and decay into a lepton and a neutrino \cite{Frigerio:2014ifa}. Likewise, the singlet in the Higgs sector can either be produced directly and decay, or mix with the SM Higgs, producing observable deviations in its properties \cite{Petraki:2007gq,Kusenko:2006rh,Merle:2013wta,Shaposhnikov:2006xi}. While tantalizing, such collider signals are also not direct evidence of frozen-in sterile neutrino dark matter. 

\textit{Cosmological Signals}:\\
Arguably the most direct probe of sterile neutrino dark matter from freeze-in consists of looking for its imprints in the early Universe. Since the $N_1$ are produced with significant energy, they effectively act as radiation in the early Universe and can contribute to the number of relativistic degrees of freedom $N_{\rm{eff}}$ during BBN or photon decoupling, which are tightly constrained from observations. The contribution to the deviation from the standard $N_{\rm{eff}}$ value can be estimated as\cite{Merle:2015oja} 
\begin{equation}
\Delta N_{\rm{eff}}=\frac{\rho- n m_{N_1}}{2\rho_{\rm{therm}}^{\rm{ferm}}},
\end{equation}
which compares the kinetic part of the sterile neutrino energy density with $\rho_{\rm{therm}}^{\rm{ferm}}$, the energy density of a perfectly relativistic sermonic species in equilibrium at the same temperature. Ref.\,\refcite{Merle:2015oja} examined the variations of this quantity over the parameter space of the extended higgs sector model \cite{Petraki:2007gq,Kusenko:2006rh,Merle:2013wta} and found sizable contributions, even exceeding the current bounds on $\Delta N_{\rm{eff}}$ from BBN in some cases. A measurable deviation in $N_{\rm{eff}}$ beyond the SM value due to sterile neutrino dark matter might therefore be within reach of future probes. Likewise, constraints from dwarf spheroidals might also hold relevance (see Ref.\,\refcite{Boyanovsky:2008nc}).


\section{Summary}
\label{sec:summary}

To conclude, the major points of this article are summarized as follows:
\begin{itemlist}
 \item Given that sterile neutrino dark matter from Dodelson-Widrow is inconsistent with X-ray and Lyman-$\alpha$ measurements, freeze-in production, where the framework automatically ensures consistency with these measurements, represents an an attractive and viable production mechanism for sterile neutrino dark matter.
 \item The framework requires a feeble coupling to some particle, generally a scalar in the thermal bath, and therefore requires physics beyond the $\nu$MSM.  Sec. \ref{sec:models} discussed scenarios where these ingredients are motivated by other physics considerations, ranging from inflation, an extended Higgs sector, additional symmetry for the sterile neutrinos and possible connections to supersymmetry, leptogenesis, and extra dimensions. Depending on the details of the model, freeze-in can be dominated by IR or UV physics. 
 \item Sterile neutrino dark matter from freeze-in carries very different features from DW production: the mixing angle with the active neutrinos can be arbitrarily small, even nonexistent; it has a colder spectrum due to production earlier in the cosmological history; and the mass need not be keV scale (models discussed in Sec.\ref{sec:models} contained MeV-GeV scale candidates). 
 \item Phenomenologically, this framework can provide large deviations to $N_{\rm{eff}}$ during BBN, the scalar involved in the freeze-in process can be direct probed at colliders, and a vanishing mass for the lightest active neutrino is a salient feature of this setup.
\end{itemlist} 

The sterile neutrino remains an attractive candidate that motivates interesting connections between the neutrino and dark matter sectors, which are two of the most active areas of contemporary research. Its incorporation within the freeze-in mechanism is still a relatively young and unexplored direction. As shown in this article, this framework is compatible with several well-defined physics ideas, and can motivate interesting connections with seemingly unrelated topics in physics, suggesting that several rich avenues remain open for exploration.  

\appendix

\section*{Acknowledgments}

The author is supported by the Department of Energy under grants DE-SC0007879 and DE-SC0011719.

\end{document}